\documentclass[twocolumn,showpacs,preprintnumbers,amsmath,amssymb]{revtex4}
\usepackage{graphicx}
\usepackage{dcolumn}
\usepackage{bm}
\begin{document}
\title{Intrinsic time-uncertainties and decoherence: \\comparison of 4 models}
\author{Lajos Di\'osi}
\email{diosi@rmki.kfki.hu}
\affiliation{Research Institute for Particle and Nuclear Physics, H-1525 Budapest 114, POB 49, Hungary}
\date{\today}
\begin{abstract}
Four models of energy decoherence are discussed from the common perspective of intrinsic 
time-uncertainty. The four authors --- Milburn, Adler, Penrose, and myself --- have four 
different approaches. The present work concentrates on their common divisors at the level of
the proposed equations rather than at the level of the interpretations. General relationships between
time-uncertainty and energy-decoherence are presented in both global and local sense. 
Global and local master equations are derived. (The local concept is favored.)
\end{abstract}

\maketitle

\section{Introduction}
Let me begin with an incomplete list of people who formulated ideas related to the
possible role of time or space-time uncertainties in the destruction of quantum
coherence. Certainly Feynman \cite{Fey62} mentioned the idea first, and a Hungarian group
\cite{Kar66,KFL82,KFL86}
developed a first vague model. From Penrose to Steve Adler and during twenty years,
many independent investigations 
\cite{Pen94,Pen96,Pen98,Dio87,Dio89,EMN89,EHNS84,Mil91,Mil03,San93,CFN94,PerStr97,PowPer00,Roy98,Adl04} 
shared the central idea and disagreed on the
motivations. In the rest of the present contribution I single out  
four authors: Penrose \cite{Pen94}, myself \cite{Dio87,Dio89}, Milburn \cite{Mil91,Mil03}, and Adler 
\cite{Adl04}. They have four different
motivations: Penrose exposes the conceptual uncertainty of location in space-time,
I attribute an ultimate uncertainty to the Newtonian gravitational field,
Milburn assumes that Planck-time is the smallest time, and Adler derives
quantum theory in the special limit of a hypothetic fundamental dynamics.
The four authors also have four different mathematical apparatuses, four different 
interpretations, metaphysics, e.t.c., but the four models have common divisors and my
present intention is to find and emphasize them. The reader shall see that 
the Milburn master equation is identical to the simplest effective equation
derived by Adler in his dynamical theory, and Penrose's equation is a special
case of my master equation.

\section{Energy decoherence}
Decoherence means the destruction of interference. When it happens to energy eigenstates,
we talk about energy decoherence. Decoherence diminishes coherent dispersion like
$\Delta E$ in case of energy decoherence. Too large
coherent dispersions mean that the system exhibits strong quantum features. On the contrary,
if the coherent dispersions are reasonably small the system looks like a classical one.
Since decoherence can diminish coherent dispersions we think that decoherence is
instrumental in the emergence of classicality in quantum systems \cite{Giuetal96}.
Decoherence of various
observables may be correlated or anti-correlated. Decoherence of the {\it local} energy will
induce decoherence of location of massive objects. There is no definite
rule as to what observable is the primary one to induce decoherence of the others and to
cause eventual classicality of the macroscopic variables. There is no apriori rule as to what
are the macroscopic observables and what is the threshold for classically small dispersions. 
{\it Yet, a consistent decoherence model, even if it is phenomenological, would make suggestions
for the primarily decohering observables and for the dynamics of their dispersions.}  
Total energy is a possible choice to start with, and it leads us to the concept of local energies 
as the primarily decohering observables. Energy decoherence has intimate relationship
with the intrinsic uncertainty of time, elucidated below in subsections \ref{Glob} and \ref{Loc}.
\begin{figure}[ht]
\includegraphics{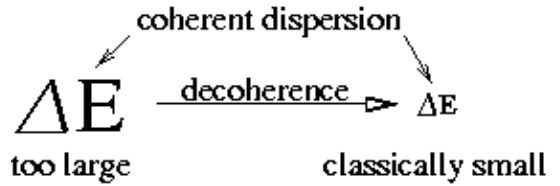}
\caption{\label{fig1} Decoherence destroys large quantum fluctuations of total energy.}
\end{figure}

Decoherence is, similarly to many other irreversible mechanisms, fueled by the 
complexity of the exact dynamics, like in the Adler theory (Sec.\ref{Adler}). The other three theories 
discussed later in Secs.\ref{Milburn},\ref{Penrose},\ref{Diosi}, 
derive decoherence from non-dynamic mechanisms. They assume inherent randomness 
of time. The typical effect is a simple exponential decay of coherence at a certain decoherence time scale
$t_D$. 
This exponential rule has been met by almost all dynamic models as well as by
the four ones singled out for the present discussion. The Milburn-Adler
decoherence-time is on the scale of the Planck-time $\tau_{Pl}$, the Di\'osi-Penrose
theory assumes non-relativistic decoherence-time. I shall abandon the complex details
of these theories and wish to compare them at the level of their ultimate phenomenological equations. 

\subsection{Global time uncertainty and decoherence}
\label{Glob}
Let me illustrate the naive phenomenology of energy decoherence emerging from time uncertainty.
For simplicity, suppose that the initial state of a quantum system is a superposition of 
only two energy eigenstates of the total Hamiltonian $H$:
\begin{equation}\label{psi}
\vert\psi\rangle = c_1\vert\varphi_1\rangle+c_2\vert\varphi_2\rangle \;\;. 
\end{equation}
The evolution of the state in time $t$ reads:
\begin{equation}\label{psit}
\vert\psi(t)\rangle = 
c_1\exp(-i\hbar^{-1}E_1t)\vert\varphi_1\rangle+c_2\exp(i\hbar^{-1}E_2t)\vert\varphi_2\rangle \;\;. 
\end{equation}
Following our purposes, we add a certain uncertainty $\delta t$ to $t$:
\begin{equation}\label{dt}
t \rightarrow t+\delta t \;\;, 
\end{equation}
and substitute it into the Eq.(\ref{psit}).
To be concrete, suppose $\delta t$ is a
Gaussian random variable of zero mean and of dispersion proportional to the mean time $t$ itself:
\begin{equation}\label{Mdtdt}
{\rm\bf M}[(\delta t)^2] = \tau t \;\;,
\end{equation}
where $\tau$ is a certain time-scale to measure the strength of time-uncertainties.
We can evaluate the evolution of the density matrix:
\begin{eqnarray}\label{rho}
&&\rho(t)\equiv{\rm\bf M}\left[\vert\psi(t)\rangle\langle\psi(t)\vert\right]=\\
         &=&\vert c_1\vert^2\vert\varphi_1\rangle\langle\varphi_1\vert+
            \vert c_2\vert^2\vert\varphi_2\rangle\langle\varphi_2\vert+\nonumber\\
         &+&\bigl\{c_1^\star c_2\exp(i\hbar^{-1}\Delta Et)
                  {\rm\bf M}\left[\exp(i\hbar^{-1}\Delta E\delta t)\right]
                   \vert\varphi_2\rangle\langle\varphi_1\vert+\nonumber\\
         &+&{~~~~~~\rm h.c.~~~~~~}\bigr\}\;\;.\nonumber
\end{eqnarray}
The stochastic mean on the r.h.s. yields:
\begin{equation}\label{Mexp}
{\rm\bf M}\left[\exp(i\hbar^{-1}\Delta E\delta t)\right] = e^{-t/t_D} \;\;, 
\end{equation}
which means that the off-diagonal term of $\rho(t)$ decays exponentially 
with the decoherence time
\begin{equation}\label{tD}
t_D = \frac{\hbar^2}{\tau}\frac{1}{(\Delta E)^2} \;\;. 
\end{equation}

It is easy to see that all off-diagonal elements of a generic initial density matrix would 
decay exponentially. The compact form of the evolution equation contains a typical
double-commutator term in addition to the standard commutator on the r.h.s. of the von
Neumann equation:
\begin{equation}\label{ME}
\frac{d\rho}{dt}=-i\hbar^{-1}[H,\rho]-\frac{1}{2}\tau\hbar^{-2}[H,[H,\rho]] \;\;.
\end{equation}
The direct proof of this master equation is simple. The reader is referred to the proof of the
more general case (\ref{MEloc}).

I anticipate that the above master equation will be obtained from slightly different concepts.
In Sec.\ref{Milburn} the time uncertainty $\delta t$ follows the Poisson statistics and the
Eq.(\ref{ME}) is obtained in the proper time-coarse-grained limit. In Sec.\ref{Adler} the
Eq.(\ref{ME}) is exactly derived from the uncertainties of the Planck-constant. The 
corresponding mathematics is trivial. Gaussian randomness $\delta t$ can mathematically be 
delegated to the randomness $\delta\hbar^{-1}$ since in the solutions of the 
Schr\"odinger---von-Neumann equation the Planck-constant and the time appear always in the 
product form $\hbar^{-1}t$. For the physics of the above two versions, see Secs.\ref{Milburn} and
\ref{Adler}. 

\subsection{Local time uncertainty and decoherence}
\label{Loc}
It is straightforward to consider a local generalization of total-energy decoherence.
We drop the concept of global uncertainty $\delta t$ of global time $t$ in favor of local 
uncertainties $\delta t_{\bf r}$ of the local time $t_{\bf r}$:
\begin{equation}\label{dtr}
t_{\bf r} \rightarrow t+\delta t_{\bf r} \;\;, 
\end{equation}
where ${\bf r}$ labels spatial cells. 
Let, by assumption, the $\delta t_{\bf r}$ be 
Gaussian random variables of zero mean and of spatial correlation proportional to the time $t$ itself:
\begin{equation}\label{Mdtrdtr}
{\rm\bf M}[\delta t_{\bf r}\delta t_{\bf r'}] = \tau_{\bf rr'}t \;\;,
\end{equation}
where $\tau$ is a certain Galileo invariant spatial correlation function.
Let us write the total Hamiltonian as the sum $H=\sum_{\bf r}H_{\bf r}$ of local ones. The solution
of the von Neumann equation takes this form:
\begin{eqnarray}\label{rholoc}
\rho(t) =\rho(0)&-&i\hbar^{-1}\sum_{\bf r}[H_{\bf r}t_{\bf r},\rho(0)]-\nonumber\\
                &-&\frac{1}{2}\hbar^{-2}\sum_{\bf r,r'}[H_{\bf r}t_{\bf r},[H_{\bf r'}t_{\bf r'},\rho(0)]]
 \;\;, 
\end{eqnarray}
plus higher order terms in $t$. Let us take the 
stochastic mean of the r.h.s., using ${\bf M}[t_{\bf r}]=t$ and 
${\bf M}[t_{\bf r}t_{\bf r'}]=\tau_{\bf rr'}t+t^2$. This leads to the following master equation
at $t\rightarrow0$:  
\begin{equation}\label{MEloc}
\frac{d\rho}{dt}=
-i\hbar^{-1}[H,\rho]-\frac{1}{2}\hbar^{-2}\sum_{\bf r,r'}\tau_{\bf rr'}[H_{\bf r},[H_{\bf r'},\rho]] \;\;.
\end{equation}
Since the matrix $\tau$ is non-negative the above class of master equations will, as expected, decohere the 
superpositions of local-energy eigenstates.

\section{Milburn: discrete time uncertainty}
\label{Milburn}
\begin{table*}
\caption{\label{tab1}Decoherence times for the Milburn master equation.}
\begin{ruledtabular}
\begin{tabular}{ll}
$\Delta E=1$eV (atomic superpositions)            & $t_D\sim10^{13}$s (irrelevant in atomic physics)\\
$\Delta E=1$GeV (high energy superpositions)      & $t_D\sim10^{-5}$s (would be irrelevant for nuclear forces)\\
$\Delta E=1$J (macroscopic superpositions)        & $t_D\sim10^{-25}$s (excludes superposition of macro-energies)\\
\end{tabular}
\end{ruledtabular}
\end{table*}
Milburn assumes discrete global time which is of the form $n\tau_{Pl}$ 
The integer $n$ is random with Poisson distribution of mean value
\begin{equation}\label{Mn}
{\rm\bf M}[n]=t/\tau_{Pl} \; \;.
\end{equation}
Hence time, too, becomes intrinsically uncertain while $t$ stands for the respective mean value
of the random time. Milburn assumes, furthermore, that the random structure of time is not 
accessible to us and we can only observe the physics averaged for the random fluctuations of time. 

It is straightforward to write down the master equation governing the evolution of the state $\rho$. 
First, we write down the change of the state in a single step of the discrete time:
\begin{equation}\label{rhotauPl}
\rho\rightarrow\exp(-i\hbar^{-1}H\tau_{Pl})\rho\exp(i\hbar^{-1}H\tau_{Pl}) \; \;.
\end{equation}
Consider an infinitesimal interval $dt$ of mean time. During it, the above step occurs with the 
infinitesimal probability $dt/\tau_{Pl}$, otherwise the state remains unchanged. Taking the
averaged change of $\rho$ during $dt$, we obtain the Milburn master equation:
\begin{equation}\label{MEMilburn}
\frac{d\rho}{dt}=\frac{1}{\tau_{Pl}}
\left[\exp(-i\hbar^{-1}H\tau_{Pl})\rho\exp(i\hbar^{-1}H\tau_{Pl})-\rho\right] \; \;.
\end{equation}
Let us expand the r.h.s. upto the first order in the Planck-time:
\begin{equation}\label{MEMilburn1}
\frac{d\rho}{dt}=-i\hbar^{-1}[H,\rho]-\frac{1}{2}\tau_{Pl}\hbar^{-2}[H,[H,\rho]] \; \;.
\end{equation}
The expansion is valid at the condition $\Delta E\tau_{Pl}\ll\hbar$. Recall that the above equation
is identical with Eq.(\ref{ME}) at the choice $\tau=\tau_{Pl}$. Its basic feature is that the 
off-diagonal elements of $\rho$ in energy representation will decay at the characteristic
decoherence time (\ref{tD}):
\begin{equation}\label{tDPl}
t_D = \frac{\hbar^2}{\tau_{Pl}}\frac{1}{(\Delta E)^2} \;\;. 
\end{equation}
This scale suggests plausible physics, at least at a quick glance. The intrinsic time-uncertainty
does practically not decohere atomic superpositions, while large energy superpositions would decay
at extreme short times (Tab.\ref{tab1}).

\section{Adler: emergent quantum mechanics}
\label{Adler}
In Adler's theory, the deepest level of dynamics is classical. The generalized coordinates are
complex $N\times N$ hermitian matrices $\{q_r\}$. Their labels $r$ can be taken as, e.g., labels of spatial
cells. The dynamics is defined by the Lagrangian:
\begin{equation}\label{Lag}
{\bf L}[\{q_r\},\{\dot q_r\}]={\rm Tr} L[\{q_r\},\{\dot q_r\}] \;\;.
\end{equation}   
It is also called trace Lagrangian, generating trace dynamics for $q_r$. Following the standard method,
Adler introduces the canonical momenta:
\begin{equation}\label{pr}
p_r=\frac{\partial{\bf L}}{\partial q_r} \;\;,
\end{equation}   
and the Hamilton function:
\begin{equation}\label{Ham}
{\bf H}={\rm Tr}\sum_r p_r\dot q_r - {\bf L}\;\;.                 
\end{equation}
If the Lagrange function $L$ is constructed from the generalized coordinates $q_r$, from the velocities 
$\dot q_r$, from complex number coefficients, and we exclude matrix coefficients then we can prove that 
the following matrix is a conserved dynamic quantity \cite{Mil97}:
\begin{equation}\label{C}
\sum_r [q_r,p_r] = {\bf\tilde C} = {\rm const.}                 
\end{equation}
The proof is straightforward. We can write:
\begin{equation}\label{Cdot}
\frac{d{\bf\tilde C}}{dt}= \frac{d}{dt}\sum_r\left[q_r,\frac{\partial{\bf L}}{\partial\dot q_r}\right]
                         = \sum_r\left(
\left[q_r,\frac{\partial{\bf L}}{\partial q_r}\right]+
\left[\dot q_r,\frac{\partial{\bf L}}{\partial\dot q_r}\right]\right)
 \;\;,                 
\end{equation}
where we applied the Euler-Lagrange equation. The r.h.s. vanishes. Indeed, from the unitary invariance of 
${\bf L}$ under, say, the infinitesimal variation $\delta q_r=i[\Lambda,q_r]$ and 
$\delta\dot q_r=i[\Lambda,\dot q_r]$ where $\Lambda$ is an
arbitrary hermitian matrix, we have:
\begin{equation}\label{dL}
0={\delta\bf L}= i{\rm Tr}\sum_r 
\left([\Lambda,q_r]\frac{\partial{\bf L}}{\partial q_r}+
      [\Lambda,\dot q_r]\frac{\partial{\bf L}}{\partial\dot q_r}
\right)
 \;\;.                 
\end{equation}
After cyclic permutations under the trace operation, we recognize the r.h.s. of Eq.(\ref{Cdot}) which
must vanish because of the arbitrariness of the hermitian matrix $\Lambda$.

The conservation rule $d{\bf\tilde C}/dt=0$ is classical but it inspires something which looks quantum 
mechanical. If we were able to prove that each term $[q_r,p_r]$ of the l.h.s. of Eq.(\ref{C}) 
is constant on its own then we would choose those constants as $i\hbar$ each. Provided furthermore that 
the choice $[q_r,p_s]=0$ is also possible for all $r\neq s$, we would obtain the structure of Heisenberg
quantum mechanics from the underlying classical matrix dynamics. Adler was able to show that the
equipartition mechanism of the classical Gibbs-statistical physics will indeed provide the desired solutions.
In a suitable {\it statistical average}, an effective theory emerges with the approximate commutation relations:
\begin{equation}\label{qpeff}
[q_r^{\rm eff},p_r^{\rm eff}]=\delta_{rs}\times{\rm const}.                 
\end{equation}
One sets ${\rm const}=i\hbar$. The classical dynamics of the effective variables turns out to be the
emergent unitary dynamics:
\begin{equation}\label{Heff}
\dot q_r^{\rm eff}= i\hbar^{-1}[H^{\rm eff},q_r^{\rm eff}],~~~~~~
\dot p_r^{\rm eff}= i\hbar^{-1}[H^{\rm eff},p_r^{\rm eff}]
 \;\;.                 
\end{equation} 
Hence, one has derived an emergent quantum-canonical structure and unitary dynamics. Adler exploits
that this structure is statistically blurred. At a closer look, the emergent quantum mechanics contains
some irreversibility. It will be easy to see the resulting master equation in Schr\"odinger representation.
We start from the effective von Neumann equation $\dot\rho=-i\hbar^{-1}[H^{\rm eff},\rho]$ and we observe that
$\hbar$ is the statistical mean of a fluctuating parameter. Let us reintroduce the fluctuations of $\hbar$:
\begin{equation}\label{dh}
\hbar^{-1}\rightarrow\hbar^{-1}+\delta\hbar^{-1} \;\;,
\end{equation}
and let us approximate them by a certain white-noise satisfying
\begin{equation}\label{dhdh}
{\bf M}[(\delta\hbar^{-1})^2]=\hbar^{-2}\tau/t \;\;,                
\end{equation} 
where the precise meaning of $\delta\hbar^{-1}$ is the fluctuation of $\hbar^{-1}$'s average over the period $t$.
Let us insert Eq.(\ref{dh}) into the von Neumann equation and let us average over $\delta\hbar^{-1}$. 
The resulting master equation reads:
\begin{equation}\label{MEAdler}
\frac{d\rho}{dt}=-i\hbar^{-1}[H^{\rm eff},\rho]-\frac{1}{2}\tau\hbar^{-2}[H^{\rm eff},[H^{\rm eff},\rho]] \; \;.
\end{equation}
This is again the master equation of energy decoherence.

In the light of numerous experimental evidences, Adler gives a detailed analysis of the possible 
parametrization, including the natural choice $\tau=\tau_{Pl}$. To date, this is perhaps the most 
complete available discussion of the observable scales of energy decoherence.

On the top of the exact Heisenberg structure (\ref{qpeff},\ref{Heff}), the trace dynamics 
is more likely to superpose local fluctuations than global ones. Therefore global fluctuations 
$\delta\hbar^{-1}$ are replaced by correlated local fluctuations $\delta\hbar^{-1}_{\bf r}$. The corresponding 
master equation will be of the form (\ref{MEloc}) with the correlation matrix yet to be specified from the 
equilibrium trace dynamics.   

\section{Penrose: uncertainty of the point-wise identity in space-times}
\label{Penrose}
``\dots when the geometries become significantly different from each other, we have no absolute means
of identifying a point in one geometry with any particular point in the other \dots, so the very idea
that one could form a superposition of the {\it matter} states within these two separate spaces becomes
profoundly obscure'' --- this is the geometer's argument against the concept of sharp geometry. 
If superpositions of states with very different mass distributions existed they should ``decay''.
\begin{figure}[ht]
\includegraphics{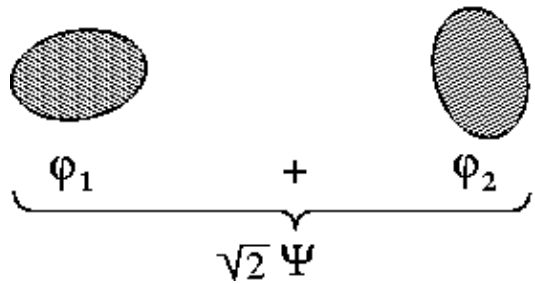}
\caption{\label{fig2} The superposition of two different mass distributions (lumps), corresponding to 
separate wave packets $\varphi_1$ and $\varphi_2$.}
\end{figure}
Penrose considers a balanced superposition of two separate wave packets representing two
different positions of a massive object (Fig.\ref{fig2}). 
If the mass $M$ is large enough, the two wave packets represent 
two very different mass distributions. Penrose postulates the following decay time $t_D$ for the 
balanced superposition of two {\it lumps}:
\begin{equation}\label{tDgrav} 
t_D=\hbar/\Delta E_{grav} \;\;,
\end{equation}
where $E_{grav}$ is the energy we must, hypothetically, supply to the system in order to separate the two 
lumps against gravitational forces: 
\begin{equation}\label{Egrav}
E_{grav}=\vert U_{11}+U_{22}-U_{12}\vert \;\;.
\end{equation}
The quantity $E_{grav}$ is the Newton self-energies $U_{11}+U_{22}$ of the two lumps minus their interaction 
energy $U_{12}$.

This is the Penrose proposal. As far as I know, there has been
no microscopic definition of the mass distributions of the lumps which enter the
calculation of the Newtonian energies. Penrose assumes the macroscopic density for the lumps and does not 
resolve it for atomic scales. This way one can calculate the decoherence time for various situations.
Since the gravitational self- or mutual energies of atomic systems are very small, the calculated 
$t_D$'s will be extremely long. The atomic superpositions will never decay in the Penrose theory.
Massive superpositions, on the other hand, would never be formed because they would decay immediately.
If, for example, one considers a rigid ball of density $\sim1$g/cm$^3$ then a plausible critical size
$R_{crit}\sim10^{-5}$cm follows, to separate microscopic from macroscopic scales. Its vague
interpretation is the following \cite{Dio87}. Let us prepare the ball of radius $R$ in a wave packet as 
narrow as $\Delta r\sim R$. Then we compare the order of dynamical time scale of the wave packet widening with
the order of decoherence time (\ref{tDgrav}). We shall see that for small balls ($R\ll R_{crit}$) the
unitary dynamics dominates while for large balls ($R\gg R_{crit}$) decoherence is quicker and wins 
over the unitary dynamics.  

Penrose also emphasizes the difference between the unitary evolution and the decay (reduction) of 
superpositions, denoted by ${\bf U}$ and ${\bf R}$ resp., in his works. While the contrary features
of ${\bf U}$ and ${\bf R}$ have been extensively discussed, the details of evolution
during reduction ${\bf R}$ have remained unspecified. Regarding the generic form of state evolution, Penrose
writes down the formal sequence  ${\bf U, R, U, R,\dots}$ while he does not construct a differential
evolution equation -- master equation -- incorporating both ${\bf U}$ and ${\bf R}$.

\section{Di\'osi: uncertainty of local time}
\label{Diosi}
In terms of local time uncertainties (\ref{dtr},\ref{Mdtrdtr}), this theory introduces the following 
universal correlation:
\begin{equation}\label{taurr}
\tau_{\bf rr'}={\rm const}\times\frac{G\hbar}{\vert{\bf r}-{\bf r'}\vert}c^{-4} \;\; ,
\end{equation}
where $G$ is the Newton constant and $c$ is the velocity of light. Let me explain the underlying 
arguments. According to the theory of general relativity, the uncertainties of local time can be
represented by the fluctuations of the $g_{00}$ component of the metric tensor. If the average
space-time is flat, which we assume for simplicity, then ${\bf M}[g_{00}]=1$ and, in Newtonian
limit, $\delta g_{00}=-2c^{-2}\Phi$ where $\Phi$ is the Newton potential. The uncertainty
of local time becomes directly related to the uncertain Newton potential $\Phi$:
\begin{equation}\label{dtrNew}
\delta t_{\bf r}\equiv\delta\int_0^t dt'g_{00}^{1/2}({\bf r},t')
\approx-c^{-2}\int_0^t dt'\Phi({\bf r},t') \;\; .
\end{equation}     
If we knew the correlation of local uncertainties of the Newtonian gravity, we could calculate
the correlation of local time uncertainties. According to heuristic calculations \cite{DioLuk87,Dio87},
the local gravitational acceleration ${\bf g}=-{\bf\nabla}\Phi$ has an inherent uncertainty, totally
uncorrelated in space and time:
\begin{equation}\label{Mgg}
{\bf M}[{\bf g}({\bf r},t){\bf g}({\bf r'},t')]
={\rm const}\times G\hbar\delta({\bf r}-{\bf r'})\delta(t-t') \;\; .
\end{equation}     
This implies the following correlation function for the Newton potential:
\begin{equation}\label{MPhiPhi}
{\bf M}[\Phi({\bf r},t)\Phi({\bf r'},t')]={\rm const}\times\frac{G\hbar}{\vert{\bf r}-{\bf r'}\vert}
                                          \delta(t-t') \;\; .
\end{equation}     
Inserting Eq.(\ref{dtrNew}) into ${\bf M}[t_{\bf r}t'_{\bf r'}]$ and using Eq.(\ref{MPhiPhi}), we arrive
at the correlation (\ref{taurr}) of local time-uncertainties. Let us emphasize that the relativistic
time-correlation function (\ref{taurr}) is equivalent with the non-relativistic gravity-correlation
(\ref{MPhiPhi}). As we shall see, the speed of light $c$ cancels from the quantum master equation. 

Now we can derive the master equation valid on the averaged space-time. We start from the general
equation (\ref{MEloc}) with the local-time correlation (\ref{taurr}). This latter is proportional to $c^4$
which makes it extremely small. In the total energy, it is only the Einstein energy which appreciates 
the time-fluctuations. We write its local decomposition as $c^2\sum_{\bf r}f({\bf r})$ where $f({\bf r})$ 
is the operator of local mass density. Then we identify $H_{\bf r}$ in the general master equation 
(\ref{MEloc}) by $H_{\bf r}=c^2\sum_{\bf r}f({\bf r})$, yielding: 
\begin{eqnarray}\label{MEDiosi}
\frac{d\rho}{dt}=&-&i\hbar^{-1}[H,\rho]\\
                 &-&\frac{G}{2}\hbar^{-1}\int\int\frac{d{\bf r}d{\bf r'}}{\vert{\bf r}-{\bf r'}\vert}
                 [f({\bf r}),[f({\bf r'}),\rho]] \;\; .\nonumber
\end{eqnarray}
This is the master equation we were looking for. [We replaced $\sum_{\bf r}$ by $\int d{\bf r}$, and
ignored the numeric factor on the r.h.s. of Eq.(\ref{MPhiPhi}).]
Observe that the $c$ has been canceled and the result is perfect nonrelativistic. This gives us the hint
that we can perform an equivalent nonrelativistic proof of the above master equation, starting
from the von Neumann equation $\dot\rho=-i\hbar^{-1}[H-\sum_{\bf r}\Phi({\bf r},t)f({\bf r}),\rho]$,
Taylor-expanding the solution upto $O(t^2)$ and calculating the stochastic mean by substituting the
gravity-correlation (\ref{MPhiPhi}).  

I am going to show that the above master equation yields exactly the Penrose decay (\ref{tDgrav},\ref{Egrav}). 
Let us start from the balanced superposition of the two massive lumps (Fig.\ref{fig2}):
\begin{equation}
\rho=\frac{1}{2}\vert\varphi_1+\varphi_2\rangle\langle\varphi_1+\varphi_2\vert \;\; .
\end{equation}
We assume that
the wave packets $\varphi_1$ and $\varphi_1$ are approximate eigenstates of the mass density operator:
\begin{equation}
f({\bf r})\vert\varphi_n\rangle=\bar f_n({\bf r})\vert\varphi_n\rangle \;\;,\;\;\;\;n=1,2;
\end{equation}
where $\bar f_1({\bf r})$ and $\bar f_2({\bf r})$ are the (c-number) mass distributions of the two
lumps, respectively. Using these functions on the r.h.s. of the master equation (\ref{MEDiosi}), 
we can write the contribution of the double-commutator term to the decay of the interference 
term between the two lumps into this form:
\begin{equation}
  \frac{d}{dt}\langle\varphi_1\vert\rho\vert\varphi_2\rangle=
  -E_{grav}\hbar^{-1}\langle\varphi_1\vert\rho\vert\varphi_2\rangle \;\;.
\end{equation}
The decay time is $t_D=\hbar/E_{grav}$ with
\begin{equation}\label{EgravDio}
E_{grav}=\int\int\frac{d{\bf r}d{\bf r'}}{\vert{\bf r}-{\bf r'}\vert}
                       [\bar f_1({\bf r})-\bar f_2({\bf r})][\bar f_1({\bf r'})-\bar f_2({\bf r'})] \;\;,
\end{equation}
i.e., it is completely identical to the Penrose proposal (\ref{tDgrav},\ref{Egrav}).

\subsection{Digression}
I would like to digress about the status of the model. An earliest criticism came from
Ghirardi, Grassi and Rimini \cite{GGG90}. The non-relativistic mass distribution $f({\bf r})$ is a basic 
ingredient of my model (as well as of Penrose's). In case of point-like constituents of position operator 
${\bf r}_k$ and mass $m_k$, the operator of mass distribution would be 
$f({\bf r})=\sum_k m_k\delta({\bf r}-{\bf r}_k)$. Newton self-energy would diverge hence my master equation 
would also diverge. I needed a short length cutoff which
I chose to be the nuclear size. The choice was naive, optimistic --- and wrong. The authors of Ref.\cite{GGG90} 
pointed out that the model can not be valid below the scale $\sim10^{-5}$cm and they suggested a higher 
cutoff.  The short-length cutoff $\sim10^{-5}$cm can most easily be implemented by the corresponding 
spatial coarse-graining of the mass density $f({\bf r})$.
Apart from this modification, the whole mathematics and physics of the model remains the same; the original
master equation needs no reformulation at all. 

The Penrose proposal avoids the cutoff-difficulty only because it uses the coarse-grained macroscopic mass 
density from the beginning, without discussing its microscopic definition.
What happens to the Penrose model if we extend it for the microscopically structured mass distributions
of the two lumps, respectively? It faces the same problem that my model did. To calculate Newtonian 
self-energies, one needs a short-length cutoff. Where should we get it from? Certainly it may be the 
phenomenological cutoff
$10^{-5}$cm imposed by Ghirardi, Grassi and Rimini \cite{GGG90} on my model. Ref.\cite{Pen98} shows Penrose's
awareness of the difficulty as well as his preliminary ideas to circumvent it. I am afraid that the proposed 
solution can still not stand on its own without defining what happens to the Newtonian interaction at short 
distances. 
  
May I clarify an important and obvious misunderstanding in Ref.\cite{Pen96}, also in some subsequent
works like, e.g., in Ref.\cite{Chr01}. The claim is that my model differs from Penrose's because I define 
the decoherence time through the inverse of the Newton interaction potential:
\begin{equation}
t_D=\hbar/\vert U_{12}\vert \;\; .
\end{equation}
This claim is obviously incorrect. 
A possible source of the misunderstanding was discovered by the late Jeeva Anandan in 1998 \cite{Ana98}: 
I had displayed a trivially mistaken version [Eq.(12) in Ref.\cite{Dio87}] of Eq.(\ref{EgravDio}), contrary 
to the otherwise correct master equation. Yet in the same work I had published the correct equation [Eq.(14)] 
as well. My longer paper \cite{Dio89} had published correct equations (4.16-17) in coincidence with 
Penrose's (\ref{tDgrav}-\ref{Egrav}), respectively.

The concrete part of the Penrose model, meaning his decay-time equation (\ref{tDgrav}-\ref{Egrav}) and its
applications, represents a special case of the concrete part of my model, i.e., my master equation 
(\ref{MEDiosi}) and its applications. The Penrose model proposes exactly the same decay times for the balanced 
superpositions of two lumps as my model does. This coincidence extends for more general superpositions as well
\cite{Ana99}. The Penrose model, however, has no dynamical equation. From the above exact coincidences
I have got the following impression. The potential dynamics, underlying the Penrose decay of massive
superpositions, can not differ from my master equation or, at least, it must build on this master equation. 
 
\section{Concluding remarks}
The purpose of this `review' was limited. It could not capture all aspects of the
chosen four models. Discussion of interpretational details was restricted to the minimum. The virtue
of my work is that I presented the four models from a common perspective which is intrinsic time-uncertainty. 
To sketch the four models on the same canvas might cause conflicts with the interpretations of the other three 
authors themselves. To reduce the risk of misunderstanding, I tried to concentrate on the concrete parts 
(equations) of the four models. I suggest the interested reader to complete his/her knowledge from the original 
references.

Let me close this work by another very incomplete list of experimental
proposals. Their diversity is spectacular. These proposals mention or suggest that the tiny 
decoherence effects, like those predicted by the above discussed theories, might (or should) be
detected in the spatial motion of a test body on a satellite \cite{KFL82}, 
in proton decay \cite{PeaSqu94},
by atom interferometer \cite{PowPer00,LJS02},
in the motion of a nano-mirror \cite{BJK99,MSPB03} or other nano-object coupled
to quantum optical devices \cite{HNDF04}, 
and in very large interferometers \cite{Ame00}. 
There are quite recent works \cite{SimJak04,Pea04} particularly concentrating on the 
experimental aspects of energy decoherence, including local energy decoherence \cite{BAI04,ABI04,Adl04a}. 
Without consistent (maybe highly phenomenological) models, the attractive new experimental options would not 
be used to target such basic quantum issue as universal decoherence. 

\section{Acknowledgments}
The author is grateful to Hans-Thomas Elze for the idea of the DICE symposiums. I thank Steve Adler for his
helpful remarks. This work was partially supported by the Hungarian OTKA Grant No. 049384.

\bibliography{apssamp}

\end{document}